\documentclass[12pt,preprint]{aastex}
\usepackage{natbib}
\usepackage{rotating}

\shorttitle{MHD Stability of ISM Phase Transition Layers}
\shortauthors{Stone \& Zweibel}

\begin{document}


\title{MHD Stability of ISM Phase Transition Layers I: 
Magnetic Field Orthogonal to Front}


\author{Jennifer M. Stone\altaffilmark{1} \& Ellen G. Zweibel\altaffilmark{1,2}}

\altaffiltext{1}{Department of Astronomy, University of Wisconsin--Madison,
    475 N. Charter Street, Madison, WI 53706}
\altaffiltext{2}{Department of Physics, University of Wisconsin--Madison,
    1150 University Avenue, Madison, WI 53706}


\begin{abstract}
We consider the scenario of a magnetic field orthogonal to a front 
separating two media of different temperatures and densities, such as cold 
and warm neutral interstellar gas, in a 2-D plane-parallel geometry. 
A linear stability analysis is performed to assess the behavior of both 
evaporation and condensation fronts when subject to incompressible, 
corrugational perturbations with wavelengths larger than the thickness of the 
front. 

We discuss the behavior of fronts in both super-Alfv\'enic and sub-Alfv\'enic 
flows. Since the propagation speed of fronts is slow in the ISM, it is the 
sub-Alfv\'enic regime that is relevant, and magnetic fields are a significant 
influence on front dynamics . In this case we find that evaporation 
fronts, which are unstable in the hydrodynamic regime, are stabilized. 
Condensation fronts are unstable, but for parameters typical of the neutral 
ISM the growth rates are so slow that steady state fronts are effectively stable.
However, the instability may become important if condensation proceeds at a 
sufficiently fast rate.

This paper is the first in a series exploring the linear and nonlinear effects 
of magnetic field strength and orientation on the corrugational instability, 
with the ultimate goal of addressing outstanding questions about small-scale 
ISM structure.
\end{abstract}




\section{Introduction}   
The interstellar medium (ISM) exhibits structure over a wide range of size
scales. In particular, so-called ``small scale structure", ranging in size 
from a few hundred kilometers up to parsec-scales, has been uncovered by a 
variety of observational techniques over the last few decades. This 
structure manifests itself in neutral, ionized and molecular form and 
appears to be over-pressured with respect to the surrounding medium. 
Small-scale atomic structure was first revealed by a VLBI HI absorption 
study towards the extragalactic source 3C 147 \citep{Die76}, which showed 
AU-scale variations in HI opacity. Other techniques to probe these variations 
include the use of optical tracers (particularly Na I D lines) against 
binary stars and globular clusters (e.g. Lauroesch et al. 1998), 
multi-epoch HI absorption measurements against pulsars (e.g. Frail et al. 
1994), and mapping of HI optical depth against continuum sources 
(e.g. Faison et al. 1998). Small-scale molecular structure is revealed by 
observations of time-variability of molecular lines seen in absorption 
against quasars (e.g. Marscher et al. 1993), and small-scale ionized 
structures are probed using pulsar scintillation (reviewed by Rickett 1990). 

At present, no single theory is able to account for all forms of small-scale
structure. The challenge is to explain the observed large ranges in pressure 
and temperature (e.g. Jenkins \& Tripp 2001) in the context of a dynamic ISM. 
The ISM is often described by a multi-phase model that includes the effects of 
supernova explosions and X-ray ionization (McKee \& Ostriker 1977). A feature 
of this paradigm is that the various phases are in approximate pressure
equilibrium. However, it is now thought that turbulence has a significant role 
in creating a more complex medium at small scales, featuring large contrasts 
in density, pressure and temperature (e.g. Kritsuk \& Norman 2002; Koyama \& 
Inutsuka 2002; Audit \& Hennebelle 2005; Gazol et al. 2005; Heitsch et al. 
2005; Mac Low et al. 2005). A major obstacle is that the fundamental nature of 
small-scale structure is not understood. The discovery of ``tiny", low column 
density clouds (e.g. Braun \& Kanekar 2005; Stanimirovi\'c \& Heiles 2005) 
suggests it could be in the form of discrete objects, but it is also argued that 
it is part of a power spectrum resulting from turbulent processes 
(e.g. Deshpande 2000). The pervasiveness of small-scale structure is also 
unknown. For example, a multi-epoch pulsar study by Frail et al. (1994) 
suggested that $\sim 15 \%$ of the cold neutral medium (CNM) is in the form of 
AU-scale structure. However, more recent pulsar work (e.g. Johnston et al. 2003; 
Stanimirovi\'c et al. 2003) shows structure to be more sporadic and argues that 
previous ``detections" were likely the result of systematic calibration errors. 
VLBA observations towards 3C 138 \citep{Br05} suggest the filling factor of 
atomic structure in the CNM is $\leq 1$ \%. Thus, it is especially complicated 
to assess the degree to which small-scale structure is universal in the ISM 
since it is difficult to detect.    

To make progress in understanding structure generation it is important to 
assess the relative roles of dyamical processes and those involving thermal 
conduction and evaporation. For example, Slavin (2007) has argued that 
evaporation is unimportant in dynamically active regions but that it may become 
dominant in quiescent regions. In this paper we examine the stability 
properties of evaporation and condensation fronts separating two media of 
different temperatures and densities. The thermal-type corrugational 
instability results from perturbations that are periodic on a surface of 
discontinuity, causing it to wrinkle. Its character has been discussed in the 
context of shock waves (D'yakov 1954; Landau \& Lifshitz 1987) and combustion 
\citep{Will64}. The instability was first applied to combustion in Type Ia 
supernovae by Bychkov \& Liberman (1995). In a study of the structure and 
stability of CNM/WNM phase transition layers Inoue et al. (2006; hereafter 
IIK06) showed that evaporation fronts are hydrodynamically unstable to 
corrugational deformations, whereas condensation fronts are hydrodynamically 
stable. It was also suggested that such perturbations could generate turbulence 
in the nonlinear regime within a dynamical time of the ISM. Three-dimensional 
hydrodynamic numerical simulations \citep{KN02} suggest that the formation of 
filaments and voids in a multiphase ISM induce turbulence that can saturate the 
nonlinear thermal instability. Our primary interest here is in the potential of 
the corruguational instability to generate structure. 

The Alfv\'en velocity in the neutral ISM is at least 10 - 100 times larger than 
the flow speed of fronts. Therefore, magnetic fields will significantly affect 
their dynamics. In this work we discuss the modifications to the hydrodynamic 
analysis for the scenario of a magnetic field orthogonal to a front, and 
concentrate on the sub-Alfv\'enic regime as the astrophysically relevant case. 
The linear effects of magnetic fields on combustion fronts in the context of 
supernovae were considered by Dursi (2004). In that work it was shown that 
super-Alfv\'enic flames are not significantly affected by a magnetic field, 
whereas in the sub-Alfv\'enic case the instability can be greatly suppressed. 
Dursi's work is equivalent to our evaporation front setup, so in this paper we 
extend the analysis to condensation fronts.   

We discuss the physical picture of fronts in \S2 and present the linear 
stability analysis in \S3. Mode analyses and dispersion relations for the 
super-Alfv\'enic and sub-Alfv\'enic regimes are presented in \S4. The physical 
mechanism of the instability is discussed in more detail in \S5. We apply our 
findings to the neutral ISM in \S6, and summarize our results and mention 
future work in \S7. In the appendices we present the derivation of the 
dispersion relation for the magnetized entropy-vortex wave and compare our 
method with that of Dursi (2004), to allow ease of comparison between the 
supernova and ISM applications. This paper is the first in a series exploring 
the linear and nonlinear effects of magnetic field strength and orientation on 
the corrugational instability. 

\section{Front Structure}
We consider the physical setup of a phase transition layer (or front) 
separating two uniform media of different densities and temperatures in a 
2-D plane-parallel geometry. These phases result from the balance of radiative 
heating and cooling due to cosmic rays and external radiation fields 
\citep{W03}. The nature of a front depends on the pressure. There exists a 
``saturation pressure" such that heating and cooling are balanced inside a 
static front (Zel'dovich \& Pikel'ner 1969; Penston \& Brown 1970). An 
evaporation front exists if the saturation pressure exceeds the external 
pressure such that there is net heating across the transition layer. In a 
condensation front the saturation pressure is exceeded by the external pressure
such that there is net cooling across the layer. The physical setup is 
illustrated in Figure \ref{BIIK}. We assume the front is located in the plane 
$x=0$ with the ambient flow parallel to the $x$-axis and select the reference 
frame so that it moves with the front. A uniform magnetic field is orthogonal 
to the front, but has no effect on its structure since it does not exert any 
force and does not modify thermal conduction in the direction of the 
temperature gradient. We assume the perturbations on each side of the front are 
incompressible, since the ambient flow speed is subsonic (IIK06). 

The thickness of a front (the Field length, $\lambda_F$) is given by
\begin{equation}
\lambda_F=\sqrt\frac{\kappa(T) T}{n^2 \cal{L}} ,
\end{equation} 
where $\kappa(T)$ is the temperature-dependent thermal conductivity and 
$n^2 \cal{L}$ is the net cooling rate \citep{BMcK90}. For example, $\lambda_F$ 
has the values $0.01$ pc and $0.1$ pc for the CNM and WNM, respectively 
(IIK06). In order to include all relevant physical processes one must consider 
perturbations with wavelengths both larger and smaller than $\lambda_F$. For 
wavelengths larger than the front thickness the transition layer is treated as 
discontinuous and perturbations of the fluid dynamical variables are important.
This allows for a straightforward analysis but neglects the effects of thermal 
conduction and thermal stability (IIK06). To remedy this one must also consider 
corrugations smaller than the front thickness for which perturbations of the 
thermodynamic variables are important. In this work we only discuss the long 
wavelength, discontinous front approximation. The roles of heating, cooling, 
and thermal conduction in front stability will be the subject of future study. 

\section{Phase Transition Layer Stability Analysis}
We perform a linear stability analysis for fronts undergoing corrugational 
deformation when subject to an orthogonal magnetic field. We choose a similar 
method to that of IIK06, rather than that used by Dursi (2004). We provide 
comments throughout this paper on the commonalities of all three approaches, 
with a more detailed comparison provided in Appendix B. It should be noted that 
Dursi's combustion front is equivalent to our evaporation front setup, in the 
sense that the downstream density is lower than the upstream density. Our work 
is thus an extension of his in providing the stability analysis for a 
condensation front.

We begin by linearizing the equations of continuity, momentum conservation, and 
induction in the incompressible approximation to obtain 
\begin{equation}
\nabla \cdot (\delta \bf v) = 0 ,
\label{cont}
\end{equation}

\begin{equation}
\frac{\partial}{\partial t}(\rho_{U,D} \delta {\bf v}) + \rho_{U,D} 
(\bf v_{\rm U,D} \cdot \nabla) \delta v + \nabla \textrm{P} = \frac{(\nabla \times 
\delta {\bf B})\times {\bf B}}{4 \pi} ,
\label{mom}
\end{equation}

\begin{equation}
\frac{\partial}{\partial t}(\delta {\bf B}) = \nabla \times 
[(\delta {\bf v} \times {\bf B}) + ({\bf v} \times \delta {\bf B})] ,
\label{bind}
\end{equation}
where subscripts ``U" and ``D" refer to upstream and downstream, respectively.
$\rho$, $\mathrm{v}$, $\mathrm{P}$, and $\mathrm {B}$ represent density, 
velocity, pressure, and magnetic field, and perturbed quantities are denoted by 
the use of ``$\delta$". We assume ideal MHD and neglect diffusivity. We take the 
front to be in the plane $\mathrm{x} = 0$ and assume perturbations in the 
({\bf k}, {\bf B}) plane of the form $\exp[i(k_xx + k_yy - \omega t)]$, so that 
unstable modes have $\mathrm{Im}(\omega)>0$. Substituting the form of the 
perturbations into equations (\ref{cont}), (\ref{mom}) and (\ref{bind}), and taking 
${\bf B} = B {\bf \hat{x}}$ (i.e. orthogonal to the front) yields the following 
equations:
\begin{equation}
k_x \delta v_x + k_y \delta v_y = 0 ,
\label{in1}
\end{equation}

\begin{equation}
\rho_{U,D}(-\omega \delta v_x + k_x v_{U,D} \delta v_x) = -k_x \delta P ,
\end{equation}

\begin{equation}
\rho_{U,D}(-\omega \delta v_y + k_x v_{U,D} \delta v_y) = 
-k_y \delta P + \frac{B}{4 \pi}(k_x \delta B_y - k_y \delta B_x) ,
\end{equation}

\begin{equation}
-\omega \delta B_x = k_y (v_{U,D} \delta B_y - B \delta v_y) ,
\end{equation}

\begin{equation}
-\omega \delta B_y = k_x (B \delta v_y - v_{U,D} \delta B_y) .
\label{in2}
\end{equation}

The perturbations are decomposed into two surface waves that are localized to the 
front, and two magnetized entropy-vortex waves that are advected by the flow. We adopt
a notation similar to that of IIK06 and denote the surface waves upstream and downstream 
of the front by $\delta v_y^{(-)}$ and  $\delta v_y^{(+)}$, with dispersion
relations given by $k_x^{(-)} = -i k_y$ and $k_x^{(+)} = i k_y$, respectively. The
magnetized entropy-vortex waves are represented by $\delta v_y^{(s-)}$ and 
$\delta v_y^{(s+)}$. Their dispersion relations are given by 
$k_x^{(s-)} = \omega / (v - v_A)$ and $k_x^{(s+)} = \omega / (v + v_A)$, where the 
Alfv\'en speed is given by $v_A = B /\sqrt{4 \pi \rho}$. The nature of these magnetized
entropy-vortex waves and the derivation of their dispersion relations are discussed
further in Appendix A. 

The directions of propagation of the magnetized entropy-vortex waves depend on whether 
the flow is super- or sub-Alfv\'enic. In the super-Alfv\'enic case no magnetized 
entropy-vortex waves can propagate upstream, so $\delta v_y^{(s-)}$ and 
$\delta v_y^{(s+)}$ are both downstream perturbations. In the sub-Alfv\'enic regime 
$\delta v_y^{(s-)}$ propagates upstream whereas $\delta v_y^{(s+)}$ is a downstream 
perturbation. In \S4 we derive separate characteristic equations for each regime.  

The $x-$component velocity, magnetic field, and pressure 
perturbations can be rewritten in terms of the $y-$component velocity 
perturbations upstream and downstream of the front, using equations (\ref{in1}) - 
(\ref{in2}). We represent perturbed quantities as sums of the surface waves and 
magnetized entropy-vortex waves: 
\begin{equation}
\delta v_y = \sum_\alpha \delta v_y^{(\alpha)}
\end{equation}

\begin{equation}
\delta v_x = - \sum_\alpha \frac{k_y}{k_x^{(\alpha)}} \delta v_y^{(\alpha)} , 
\label{up}
\end{equation}

\begin{equation}
\delta B_y = \sum_\alpha \frac{k_x^{(\alpha)} B}{(v_{U,D} k_x^{(\alpha)} - \omega)}
\delta v_y^{(\alpha)} ,
\label{yind}
\end{equation}

\begin{equation}
\delta B_x = \sum_\alpha \frac{- k_y B}{(v_{U,D} k_x^{(\alpha)} - \omega)}
\delta v_y^{(\alpha)} ,
\end{equation}

\begin{equation}
\delta P = \sum_\alpha \Bigg[\frac{B^2 (k_x^{(\alpha)2} + k_y^2)}
{4 \pi k_y (v_{U,D} k_x^{(\alpha)} - 
\omega)}  - \frac{\rho_{U,D}}{k_y}(v_{U,D} k_x^{(\alpha)} - \omega)\Bigg] 
\delta v_y^{(\alpha)} ,
\label{down}
\end{equation}
where $\alpha$ corresponds to the different types of perturbation: (-),(+),(s-), and 
(s+). These quantities are computed upstream and downstream of the front, with the
appropriate perturbations substituted for $\delta v_y^{(\alpha)}$, depending on whether
the flow is super-Alfv\'enic or sub-Alfv\'enic.

To proceed, we require the jump conditions across the front. The normal and tangential 
unit vectors are given by:
\begin{equation}
\hat{n} = \hat{x} - \hat{y} \frac{\partial}{\partial y} \delta x_f , \hspace{2 cm}
\hat{t} = \hat{y} + \hat{x} \frac{\partial}{\partial y} \delta x_f ,
\end{equation}
where $\delta x_f$ is the front displacement.
These are used to compute the velocity perturbations normal and tangential to 
the front:
\begin{equation}
\delta v_n = \delta v_x + i \omega \delta x_f, \hspace{2 cm} \delta v_t = 
\delta v_y + i k_y v_{U,D} \delta x_f ,
\label{jump}
\end{equation} 
(IIK06). The expression for $\delta v_n$ includes the motion of the front 
itself. 

The jump condition due to the continuity equation is given by 
\begin{equation}
[\rho \delta v_n] = 0 ,
\end{equation}
where $[\Psi]$ represents the jump in quantity $\Psi$ across the front. Following 
IIK06, we modify this condition by assuming that the structure of the 
front is unchanged by a perturbation such that:
\begin{equation}
\rho_U \delta v_{n, U} = \rho_D \delta v_{n, D} = 0 .
\label{j1}
\end{equation}
This is equivalent to the condition used by Dursi (2004) that both sides of 
the front travel with the same velocity ($[\delta v_x] = 0$ in our notation). 

The ${\bf \nabla \cdot B}=0$ condition implies that the normal component of the 
magnetic field perturbation is conserved across the front, thus:
\begin{equation}
[\delta B_x] = 0
\end{equation}
Linearization of the normal component of the momentum flux conservation 
equation yields 
\begin{equation}
2 \rho \delta v_n v_n + \delta P - \frac{B_n \delta B_n}{4 \pi} = 0 .
\end{equation}
By invoking the continuity of $\delta B_n$ the corresponding jump condition is given by: 
\begin{equation}
\Big[2 \rho v \delta v_n + \delta P \Big] = 0 .
\end{equation}
Linearization of the tangential component of the momentum flux conservation 
equation yields the jump condition 
\begin{equation}
\Bigg[\rho v \delta v_t - \frac{B_n \delta B_t}{4 \pi} \Bigg] = 0 ,
\label{tmom}
\end{equation}
where $\delta B_t = i k_y B \delta x_f + \delta B_y$. 
By substituting the expressions for the velocity, pressure and magnetic field 
perturbations, equations (\ref{up}) - (\ref{down}), into the jump conditions, equations 
(\ref{j1}) - (\ref{tmom}), we obtain the characteristic equation of the system. This is 
written in the form:
\begin{equation}
\bf{M} {\bf \eta} = 0 ,
\end{equation}
where the matrix $\bf M$ is derived from the jump conditions and
\begin{equation}
{\bf \eta} = (\delta x_f, \delta v_y^{(-)}, \delta v_y^{(+)} , 
\delta v_y^{(s-)}, \delta v_y^{(s+)})^T ,
\end{equation}
To obtain the dispersion relation governing the perturbations we set the determinant of 
$\bf M$ equal to zero and solve for $\omega$. 

\section{Mode Analysis}
In this section we provide the characteristic equations for super-Alfv\'enic and 
sub-Alfv\'enic fronts, in terms of the density jump across the front, $r_d = 
\rho_D/\rho_U$, and the ratio of the downstream Alfv\'en speed to the downstream flow 
speed, $A = v_{A D}/v_D$. Super-Alfv\'enic fronts have $v_U > v_{A U}$ and $v_D > 
v_{A D}$, i.e. $\sqrt r_d > A$ on the upstream side and $1 > A$ on the downstream side. 
Sub-Alfv\'enic fronts have $v_U < v_{A U}$ and $v_D < v_{A D}$, i.e. $\sqrt r_d < A$ on 
the upstream side and $1 < A$ on the dowstream side. A third regime, called ``trans-Alfv\'enic"
\citep{D04}, may be identified in which the flow speed exceeds the Alfv\'en speed on one side, 
but is less than the Alfv\'en speed on the other side, i.e. A takes a value between $\sqrt r_d$ 
and unity ($\sqrt r_d < A < 1$ for evaporation, $1 < A < \sqrt r_d$ for condensation). 
The front behavior in this regime cannot be found using the methods discussed in this work, as 
it is possible for more waves to propagate than there are boundary conditions constraining them, 
and thus no unique solution exists for the initial-value problem \citep{D04}. We argue in \S6 that 
the sub-Alfv\'enic regime is of most relevance to the interstellar medium, so do not provide any 
further discussion of the trans-Alfv\'enic regime. The stability properties for both front types 
in the various regimes are summarized in Table 1.

\subsection{Super-Alfv\'enic Fronts}
If a front is super-Alfv\'enic no entropy-vortex waves can propagate upstream, so the 
total upstream y-velocity perturbation is simply given by the upstream surface wave:
\begin{equation}
\delta v_y = \delta v_y^{(-)}
\end{equation} 
Downstream, the total y-velocity perturbation is given by:
\begin{equation}
\delta v_y = \delta v_y^{(+)} + \delta v_y^{(s-)} + \delta v_y^{(s+)} .
\end{equation}
The velocity, magnetic field, and pressure perturbations on each side of the
front are written in terms of these total y-velocity perturbations, and
substituted into the jump conditions.
Collecting terms in each perturbation, the matrix, {\bf M}, of the characteristic 
equation for the super-Alfv\'enic case is given by:
\begin{equation}
{\bf M}=\left( \begin{array}{ccccc}
\omega & -1 & 0 & 0 & 0 \\
\omega & 0 & 1 & \frac{ i \omega_D}{\omega} (1-A) & 
\frac{ i \omega_D}{\omega}(1+A) \\
0 & 1+\frac{i \omega}{r_d \omega_D} & 1-\frac{i \omega}{\omega_D} & 
\frac{i \omega_D}{\omega}(1-A)(2-A) & \frac{i \omega_D}{\omega}(1+A)(2+A) \\
i \omega_D (r_d-1) & 1-\frac{i \omega_D A^2}{\omega+i r_d \omega_D} & 
-1+\frac{i \omega_D A^2}{i \omega_D-\omega} & -1+A & -1-A \\
0 & \frac{-\omega}{i r_d \omega_D + \omega} & 
\frac{-\omega}{i \omega_D -\omega} & 1 - \frac{1}{A} & 1 + \frac{1}{A} \\
\end{array} \right) \ ,
\end{equation} 
where $\omega_D = k_y v_D$.

Taking the determinant of this matrix yields the dispersion relation:
\begin{equation}
\frac{[(A^2 -1)\omega_D^2 - 2 i \omega_D \omega + \omega^2][-i r_d^3
\omega_D^3 + \omega^3 + i r_d^2 \omega_D (\omega_D^2 + 3 i \omega_D \omega +
\omega^2)]}{A r_d \omega_D (r_d \omega_D - i \omega)(\omega_D + i \omega)
\omega} = 0
\end{equation}
This can be simplified by removing two ``trivial" modes that do not cause a
displacement of the front. These modes are given by $\omega = i(1 \pm A) 
\omega_D$. Factoring these out, the dispersion relation can be written as:
\begin{equation}
\omega^3 + i r_d \omega_D \frac{(r_d+3)}{(r_d+1)} \omega^2 + \frac{r_d
\omega_D^2}{(r_d+1)} (1+2A^2 - 3 r_d) \omega + i r_d^2 \omega_D^3
\frac{(1-r_d)}{(1+r_d)} = 0 .
\label{super}
\end{equation}
In the limit $A \rightarrow0$ the hydrodynamic dispersion relation of IIK06 is
recovered. The super-Alfv\'enic stability behavior is similar to that of the 
hydrodynamic case; condensation fronts remain stable while an unstable mode is 
present in evaporation fronts. As found by Dursi (2004), the growth rate of 
this unstable mode is slightly enhanced. We discuss the physical origin of this
behavior in \S5.

\subsection{Sub-Alfv\'enic Fronts}
For sub-Alfv\'enic fronts the upstream y-velocity perturbation is composed of a
surface wave and entropy-vortex wave:
\begin{equation}
\delta v_y = \delta v_y^{(-)} + \delta v_y^{(s-)} ,
\end{equation}
and the downstream y-velocity perturbation is given by:
\begin{equation}
\delta v_y = \delta v_y^{(+)} + \delta v_y^{(s+)} .
\end{equation} 
The matrix, {\bf M}, of the characteristic equation for the sub-Alfv\'enic case is:
\begin{equation}
{\bf M}=\left( \begin{array}{ccccc}
\omega & -1 & 0 & \frac{i \omega_D}{\omega} \sqrt r_d (\sqrt r_d -A) & 0 \\
\omega & 0 & 1 & 0 & \frac{ i \omega_D}{\omega}(1+A) \\
0 & 1+\frac{i \omega}{r_d \omega_D} & 1-\frac{i \omega}{\omega_D} & 
\frac{i \omega_D}{\omega}\sqrt r_d(A-\sqrt r_d)(2-\frac{A}{\sqrt r_d}) & 
\frac{i \omega_D}{\omega}(1+A)(2+A) \\
i \omega_D (r_d-1) & 1-\frac{i \omega_D A^2}{\omega+i r_d \omega_D} & 
-1+\frac{i \omega_D A^2}{i \omega_D-\omega} & 1-\frac{\sqrt A}{r_d} & -1-A \\
0 & \frac{-\omega}{i r_d \omega_D + \omega} & 
\frac{-\omega}{i \omega_D -\omega} & \frac{\sqrt r_d}{A} - 1 & 
1 + \frac{1}{A} \\
\end{array} \right) \ .
\end{equation}
There are two trivial modes, given by $\omega=(1+A)\omega_D$ and 
$\omega=i(\sqrt r_d A - r_d) \omega_D$, which do not displace the front. 
Factoring these out, the dispersion relation may be written as:
\begin{equation}
\omega^3 - \frac{i \sqrt r_d \omega_D}{1+r_d}[1-2A(1+\sqrt r_d) - r_d] \omega^2
- \frac{r_d \omega_D^2}{1+r_d}[1+2A(A-1) + 2 \sqrt r_d (A-2) + r_d] \omega - i
r_d^{3/2} \omega_D^3 \frac{1-r_d}{1+r_d} = 0
\label{sub}
\end{equation} 

\begin{figure}
\epsscale{0.8}
\plotone{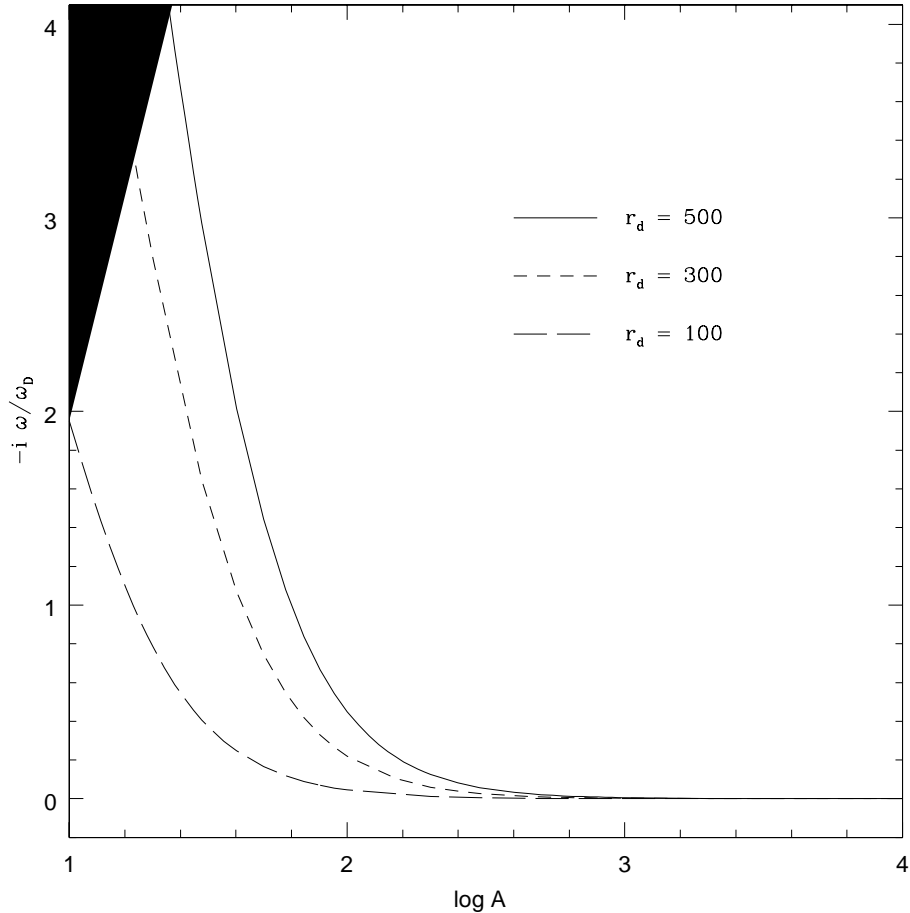}
\caption{\small Growth rate of the unstable mode in condensation fronts of various
$r_d$ as a function of A for the sub-Alfv\'enic regime. The growth rate is given in 
units of $\omega_D$. As the magnetic field strength is increased, the role of the 
density jump across the front in driving the instability becomes much less significant. 
For $A > 1000$, as expected in the neutral ISM, the growth rate is very slow, owing to 
the rigidity of the magnetic field. The blacked-out area in the top left corner covers 
the region of parameter space ($A < \sqrt r_d$) that is outside of the sub-Alfv\'enic regime.    
\label{growth}}
\end{figure}

The stability properties are the opposite of what is obtained in the 
super-Alfv\'enic case: the evaporation front is stabilized, but an unstable mode 
exists in condensation fronts, the growth rate of which decreases with increasing 
magnetic field strength. For very large magnetic field strength, as is expected for 
the ISM (see \S6), the growth rate is approximately:
\begin{equation}
\omega \approx \frac{i r_d^{1/2} (r_d-1)}{2 A^2} \omega_D
\end{equation} 
This can be written in more physical terms if we assume that material crosses the front
in a time of order the cooling time, given by $t_c=5kT/2n\Lambda$, where $n \Lambda$ is 
the cooling rate. For our calculations we use the cooling function of IIK06, in which 
$n \Lambda = 2.0 \times 10^{-26}$ ergs s$^{-1}$ corresponds to thermal equilibrium. 
Assuming the most unstable scale is twice the Field length (as is found by IIK06), the 
growth rate is given by: 
\begin{equation}
\omega \approx \frac{i r_d^{1/2} (r_d-1)}{2 A^2} \frac{2 \pi n \Lambda}{5 k_B T}
\label{cool}
\end{equation}
The growth rates of the unstable condensation mode in the sub-Alfv\'enic regime are 
plotted as a function of A for various $r_d$ in Figure \ref{growth}. The role of the 
density jump across the front in driving the instability decreases in importance with 
increasing A. 

We have also examined the eigenfunctions of the sub-Alfv\'enic matrix to 
assess the contributions of the various perturbations. The amplitudes of the
entropy-vortex waves relative to the surface waves are of order $A^{-3}$. Thus, the
instability produces very little disturbance upstream or downstream and remains 
localized to the front.

\section{Physical Mechanism}   
To investigate the mechanism of the corrugational instability we begin with the simple 
case of a front in the hydrodynamic limit, in which evaporation fronts are unstable 
whereas condensation fronts are stable. Since the normal component of the velocity 
perturbation is zero on both sides of the front, we use equation (\ref{jump}) to write:
\begin{equation}
\delta v_x = -i \omega \delta x_f.
\end{equation} 
If $i \omega$ is a positive, real number, as is the case for evaporation 
fronts, the x-velocity perturbation will reinforce the displacement of the front and 
continue to drive the instability. In condensation fronts $\delta v_x$ opposes the 
displacement and thus stabilizes the front. Since all growing modes have positive 
imaginary $\omega$, this argument is true in general for all unstable fronts, whether 
super-Alfv\'enic or sub-Alfv\'enic.

\begin{figure}
\epsscale{1.0}
\plotone{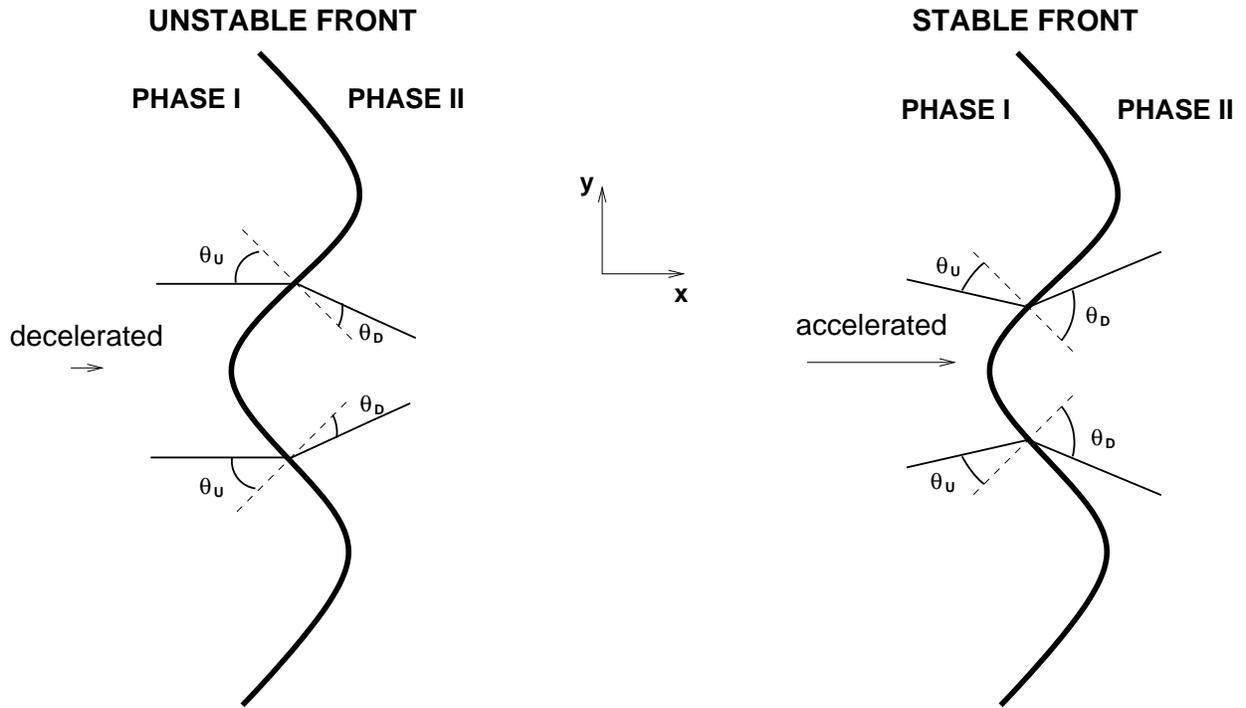}
\caption{\small Upstream and downstream streamline angles for stable and unstable 
fronts. The front is represented by the thickest line and the dotted line is normal to 
it. $\theta_U$ and $\theta_D$ represent the upstream and downstream angles, 
respectively. The thin solid line represents a given streamline. In an unstable front 
the downstream angle is smaller than the upstream angle, so streamlines are bent towards 
the normal. This accelerates the front, enabling it to propagate further ahead and drive 
the instability. In a stable front the streamlines diverge and the corrugation is smoothed 
out. 
\label{streamIIK}}
\end{figure}

Streamline curvature can also be used as a diagnostic of stability, as 
has been done for flame fronts in Type Ia supernovae \citep{D04}. In that work it was
argued that in regions where the streamlines ``fan out", the local flow velocity into the 
front is reduced and the front can propagate further ahead. We demonstrate this 
analytically, by calculating the angle formed between a given streamline and the front 
normal, to first order in the perturbation amplitude, using:
\begin{equation}
\sin \theta_{U,D} = \frac{\hat{n} \times {\bf v_{U,D}}}{|v|} = \frac{\delta v_t}{|v|}
\end{equation}
In the hydrodynamic case, $\delta v_t$ is continuous across the front, so we have 
$\sin{\theta_U}=\delta v_t/v_U$ and $\sin\theta_D=\delta v_t/v_D$. Evaporation fronts 
have $v_D > v_U$, so the downstream angle is smaller than the upstream angle so the 
streamlines bend towards the front normal. In the case of a condensation front 
$v_U > v_D$ so the streamlines bend away from the normal. This is illustrated in Figure 
\ref{streamIIK}. In regions where the streamlines diverge the $x$-velocity 
perturbation is negative ($\delta v_{x}<0$) and so the amplitude of the corrugation 
is enhanced.

In the case of a magnetic field orthogonal to the front, this analysis is more 
complicated, since the condition on $\delta v_t$ is changed. We derive it using the 
y-component of the linearized induction equation (\ref{yind}) and the tangential 
component of the linearized momentum equation (\ref{tmom}). We assume 
$\omega << k_x v_{U,D}$, which can be justified \textit{a posteriori}, and write the 
tangential momentum balance condition as:
\begin{equation}
\left(1-\frac{A^2}{r_d}\right)\delta v_{t U}=\left(1-A^2\right)\delta v_{t D}.
\label{tanmom}
\end{equation}
Upstream and downstream angles may be compared by writing equation (\ref{tanmom}) as:
\begin{equation}
\frac{\delta v_{t U}}{v_U} = \frac{A^2 -1}{\frac{A^2}{r_d} - 1} 
\frac{1}{r_d}\frac{\delta v_{t D}}{v_D} ,
\end{equation}   
such that
\begin{equation}
\sin \theta_U = \sin \theta_D \frac{A^2 -1}{A^2 - r_d} .
\label{bang}
\end{equation}
In the super-Alfv\'enic regime we consider A to be small compared to both 1 and $r_d$, 
so equation (\ref{bang}) becomes $\sin \theta_U \approx r_d^{-1} \sin \theta_D (1 - A^2 
+ A^2/r_d)$. If $r_d < 1$, the change in streamline angle across the front is increased 
by the magnetic field, so the instability is enhanced in evaporation fronts, as noted in
\S4.1.. For $r_d > 1$ the angle decreases across the front so condensation fronts are 
stable. In the sub-Alfv\'enic regime, in which $A >> 1$ we obtain $\sin \theta_U \approx 
\sin \theta_D [1 + (r_d - 1)/A^2]$, which is consistent with the stabilization of 
evaporation fronts and the slight destabilization of condensation fronts. Thus, the 
streamline picture explains the regimes of stability and instability found from the 
dispersion relations given by equations (\ref{super}) and (\ref{sub}).  

\section{Application to the ISM}
The structure of fronts is calculated in IIK06 by solving the energy equation while 
assuming thermal equilibrium at infinity. In future work we will study the effects of 
magnetic fields on front structure, but for now we simply use their results to examine 
the growth rates of unstable modes for parameters appropriate to the neutral ISM. The 
magnetic field strength parameter, $A$, can be written in terms of the upstream gas 
conditions as:
\begin{equation}
A = \frac{v_{A D}}{v_D} = \frac{B}{v_U}\sqrt{\frac{R T_U r_d}{4 \pi \mu P_U}} ,
\label{A}
\end{equation}
where $R$ is the molar gas constant, $\mu$ is the mean molecular weight (assumed to be
unity), $T_U$ is the temperature in the upstream gas, and $P_U$ is the upstream pressure. 
The parameters corresponding to examples of steady state evaporation and condensation fronts, 
derived from the front calculations of IIK06, are presented in Table 2. Both types of front are 
sub-Alfv\'enic, with their dynamics dominated by the magnetic field. In this 
regime evaporation fronts are stable, so we now calculate the growth rate of the unstable 
condensation mode. Using equation (\ref{cool}) and the parameters given in Table 2, we find 
a growth timescale $t_{grow} = 1/\omega \sim 1.8 \times 10^{18} \textrm{ s}$. This would 
imply that the corrugational instability is essentially stabilized in the neutral ISM.

However, if condensation proceeds more rapidly the instability may grow within a dynamical
time of the ISM. Inoue et al. (2007) argue that the condensation speed during the growth of 
a thermal instability that can generate clouds is of order $1$ km s$^{-1}$. If we use this velocity 
with the other parameters in Table 2 we obtain A $\sim 360$. Assuming the most unstable scale is of 
order $0.1$ pc we find a growth timescale $t_{grow} \sim 0.6$ Myr. This would imply that the 
corrugational instability has a significant effect on cloud dynamics.  

Since we have neglected the thickness of the front in this analysis we cannot provide further 
comments on the physical scale of the instability or its growth time. The full significance of the 
unstable condensation mode and the actual speed of the condensation process will only become 
apparent through completing the finite front analysis and nonlinear simulations.     

\section{Summary and Conclusions}

We have shown that for the case of a magnetic field orthogonal to a phase 
transition layer unstable modes can exist in both evaporation and condensation fronts, 
depending on whether the flow is super-Alfv\'enic or sub-Alfv\'enic. This is in contrast 
to the hydrodynamic case in which an unstable mode only occurs in evaporation fronts. We
demonstrated this by deriving linear dispersion relations, given by equations (\ref{super}) and
(\ref{sub}), and from an analysis of streamline geometry near the corrugated front. 

In the super-Alfv\'enic case the instability in evaporation fronts is enhanced from the 
hydrodynamic case, which is apparent from the increase in streamline bending. 
In the sub-Alfv\'enic case evaporation fronts are stabilized and it is the 
condensation front that is unstable. A simple formula for the maximum growth rate of the 
condensation front instability is given by equation (\ref{cool}), from which it is shown that 
under conditions appropriate to steady fronts in the neutral ISM the growth time is sufficiently 
long that such systems are effectively stable. We make the simple argument that the corrugational 
instability is suppressed by the rigidity of the magnetic field. However, if condensation proceeds 
sufficiently rapidly the growth timescale may become less than the dynamical time, such that the 
instability is important for cloud formation dynamics. 

Forthcoming papers in this series will consider perturbations smaller than the front 
thickness for both orthogonal and tangential magnetic field orientations. The full 
significance of the unstable modes will only be revealed through studies of their 
nonlinear behavior. Numerical simulations will be performed to assess the nonlinear 
characteristics of the instability and its potential for generating small-scale 
structure in the ISM. 

\acknowledgments
We acknowledge support from NASA ATP Grant NNGO5IGO9G, and NSF Grants AST-0507367 and 
PHY-0215581. We are very grateful to our anonymous referees whose recommendations 
greatly improved this paper. This work has also benefited from useful discussions with 
F. Heitsch, K. M. Hess, A. S. Hill, S.-I. Inutsuka, N. A. Murphy and S. Stanimirovi\'c.  

\newpage
\appendix
\renewcommand{\theequation}{A-\arabic{equation}}
\begin{center}
{\bf APPENDIX A}
\end{center}
{\bf Derivation of the Magnetized Entropy-Vortex Wave Dispersion Relation}

The magnetized entropy-vortex wave is a vortical perturbation that bends 
magnetic field lines and generates an Alfv\'enic response. The derivation of its 
dispersion relation is provided here for the interested reader. We seek a 
perturbation downstream of the interface satisfying $\nabla \cdot \delta v = 0$ 
while including the effects of magnetic fields. The derivation is modeled after 
that for the unmagnetized case \citep{LL87}.  

We begin by writing linearized equations for the entropy, pressure and magnetic 
field perturbations: 
\begin{equation}
i(k_x v - \omega) \delta S = 0 ,
\end{equation}

\begin{equation}
i(k_x v - \omega) \delta P = -\rho c_s^2 {\bf \nabla \cdot} \delta {\bf v} ,
\end{equation}

\begin{equation}
i(k_x v - \omega) \delta B_x = -i k_y B \delta v_y ,
\end{equation}

\begin{equation}
i(k_x v - \omega) \delta B_y = i k_x B \delta v_y .
\end{equation}
 
The velocity perturbations may be found using the $x-$ and $y-$ components of the 
linearized momentum equation: 
\begin{equation}
i(k_x v - \omega) \delta v_x = -i k_x \frac{\delta P}{\rho} ,
\end{equation}    

\begin{equation}
i(k_x v - \omega) \delta v_y = -i k_y \frac{\delta P}{\rho} + \frac{i B}{4 \pi 
\rho}(k_x \delta B_y - k_y \delta B_x) .
\end{equation}  

The energy and linearized induction equations (A-2,3,4) can be used to rewrite these 
components as:
\begin{equation}
\delta v_x = k_x c_s^2 \frac{(k_x \delta v_x + k_y \delta v_y)}{(k_x v - 
\omega)^2} ,
\label{ind}
\end{equation}

\begin{equation}
\delta v_y = k_y c_s^2 \frac{(k_x \delta v_x + k_y \delta v_y)}{(k_x v - 
\omega)^2} + \frac{k^2 v_A^2 \delta v_y}{(k_x v - \omega)^2} .
\label{nrg}
\end{equation}

Multipying (\ref{ind}) by $k_x$ and (\ref{nrg}) by $k_y$ and rearranging yields:
\begin{equation}
\Bigg[1 - \frac{k^2 c_s^2}{(k_x v - \omega)^2}\Bigg](k_x \delta v_x + k_y 
\delta v_y) = \frac{k^2 v_A^2}{(k_x v - \omega} k_y \delta v_y .
\end{equation}   
The only solution that gives ${\bf \nabla \cdot} \delta {\bf v} = 0$  is $k^2 = 0$, 
i.e.\ $k_x = \pm i k_y$, corresponding to exponentially decaying perturbations. 
These are surface waves. 

We proceed with the derivation by taking the linearized equations of motion and 
induction: 
\begin{equation}
\frac{\partial}{\partial t}\delta {\bf v} = -\frac{1}{\rho}{\bf \nabla \cdot} 
\Bigg[\delta P + \frac{{\bf B \cdot} \delta {\bf B}}{4 \pi}\Bigg] + 
\frac{{\bf B \cdot \nabla} \delta {\bf B}}{4 \pi \rho} ,
\end{equation}

\begin{equation}
\frac{\partial}{\partial t} \delta {\bf B} = {\bf \nabla \times} (\delta {\bf v
\times B}) ,
\end{equation}
and computing the curl of these to obtain: 
\begin{equation}
\frac{\partial}{\partial t} ({\bf \nabla \times} \delta {\bf v}) = \frac{{\bf B 
\cdot \nabla}}{4 \pi \rho}({\bf \nabla \times} \delta {\bf B}) ,
\end{equation}

\begin{equation}
\frac{\partial}{\partial t} ({\bf \nabla  \times} \delta {\bf B}) = 
{\bf \nabla \nabla \cdot} (\delta {\bf v \times B}) - 
\nabla^2 (\delta {\bf v \times B}) .
\end{equation}

The $z-$component of the curl of the induction equation is:
\begin{equation}
-i \omega (\nabla \times \delta {\bf B})_z = \nabla^2 (B \delta v_y) = -k^2 B 
\delta v_y ,
\end{equation}
and the vorticity is:
\begin{equation}
(\nabla \times \delta {\bf v})_z = \frac{\partial}{\partial x} \delta v_y - 
\frac{\partial}{\partial y} \delta v_x = \frac{i k^2}{k_x} \delta v_y .
\end{equation}
Combining these yields: 
\begin{equation}
-i \omega \Bigg(\frac{i k^2}{k_x} \delta v_y \Bigg) = \frac{i k_x B}
{4 \pi \rho}\Bigg(\frac{-i k^2 B}{\omega} \delta v_y \Bigg) .
\end{equation}

The solutions to the above are $k^2 = 0$, or $\omega ^2 = k_x^2 v_A^2$, so in a 
moving medium the dispersion relation of the magnetized entropy-vortex wave is 
given by: 
\begin{equation}
(\omega - k_x v)^2 = k_x^2 v_A^2,
\end{equation}
where
\begin{equation}
k_x = \frac{\omega}{v\pm v_A} .
\end{equation}
This can be compared to the dispersion relation for the unmagnetized 
entropy-vortex wave, which is simply $k_x=\frac{\omega}{v}$.

\newpage
\
\begin{center}
{\bf Appendix B}
\end{center}
{\bf Comparison with Corrugational Instability in Burning Fronts (Dursi 2004)}

\begin{figure}
\epsscale{1.0}
\plotone{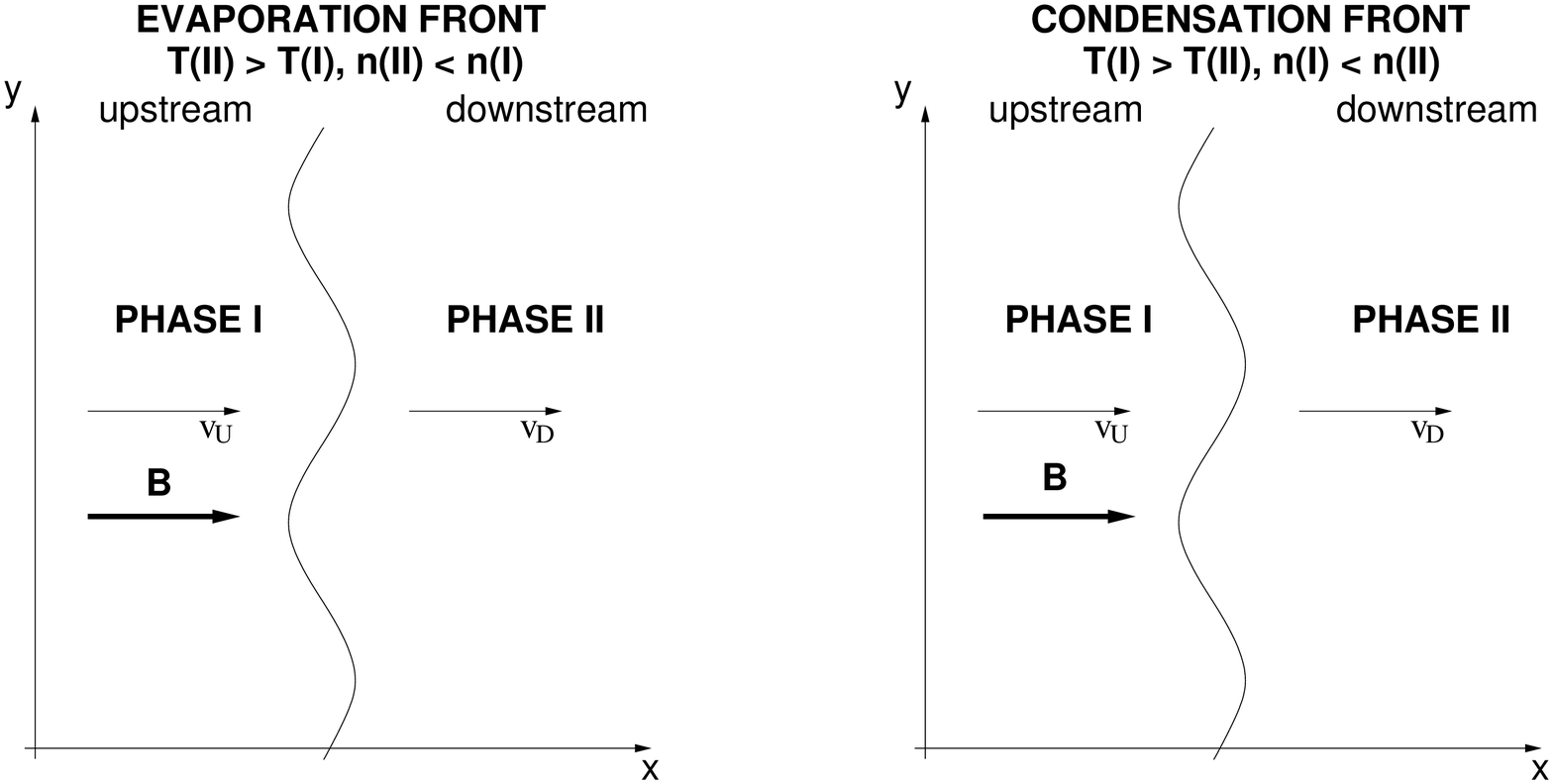}
\caption{\small Physical setup of ISM phase transition layers undergoing 
corrugational deformation in a plane-parallel geometry. The regions to the left 
and right of fronts are identified as upstream and downstream, respectively. The
phases are uniform and have different temperatures and densities. $v_U$ and $v_D$ are 
the upstream and downstream flow velocities, respectively. In an evaporation front 
there is net heating across the transition layer and in a condensation front there 
is net cooling. A magnetic field is orthogonal to the front. 
\label{BIIK}}
\end{figure}

The effect of a perpendicular magnetic field on the linear instability of flame fronts 
in Type Ia supernovae was demonstrated by Dursi (2004). Here, we compare his method with 
ours and summarize his findings, to allow comparison between the ISM and supernova 
applications. The physical setup of the supernova flame front is similar to that shown by 
the left side diagram in Figure \ref{BIIK}; one should consider ``Phase I" to be unburned 
fuel and ``Phase II" to be burned ash. Dursi's x-z coordinate system is obtained by 
rotating ours 90 degrees about the y-axis and then 90 degrees about the z-axis. The 
magnetic field is orthogonal to the flame. 

Dursi assumes incompressibility and that the reaction driving a supernova flame front is 
sufficiently exothermic such that the density of the burned material is less than that of 
the incoming fuel. Thus, his scenario is equivalent to our evaporation front setup. To 
allow ease of comparison we rewrite his notation in terms of ours. The density ratio of 
the unburned to burned material is given by:
\begin{equation}
\alpha = \frac{\rho_u}{\rho_b} = \frac{1}{r_d}
\end{equation}
The strength of the magnetic field is represented by the ratio of the Alfv\'en speed in 
the unburned material to the flow speed:
\begin{equation}
\overline a_u = \frac{a_u}{W_u} = \frac{A}{\sqrt r_d}
\end{equation} 
The growth rate itself is given by:
\begin{equation}
\overline n = \frac{n}{k W_u} = \frac{-i \omega}{r_d \omega_D}
\end{equation}                                     

It can be shown that the jump conditions we have derived and those employed by Dursi are 
the same. The only differences are our assumption that the front structure is unchanged 
by perturbations and Dursi's additional condition that each side of a front travels at 
the same velocity ($[\delta v_x] = 0$). Although this is not a condition that we directly 
impose it is still satisfied by our eigenfunctions. The dispersion relations we present 
are identical to those that are derived in Dursi (2004) and can be recovered by 
substitution of the relations given above. We have extended Dursi's work by giving full 
consideration of the behavior of condensation fronts and applying our results to the 
neutral ISM.



\clearpage

\clearpage

\clearpage

\begin{deluxetable}{ccccc}
\tabletypesize{\small}
\tablecaption{Summary of Front Stability Properties}
\tablewidth{0pt}
\tablehead{
  \colhead{Front Type} & \colhead{Hydrodynamic} &
  \colhead{Super-Alfv\'enic} & \colhead{Trans-Alfv\'enic\tablenotemark{a}} &
  \colhead{Sub-Alfv\'enic}
}
\startdata
{\bf Evaporation} & unstable & unstable & unknown & stable \\
{\bf Condensation} & stable   & stable & unknown & unstable \\
\tableline
\enddata  
\tablenotetext{a}{It is beyond the scope of this work to ascertain front stability 
in this regime. Non-linear simulations will be required.}
\end{deluxetable}

\clearpage

\begin{deluxetable}{ccccccc}
\tabletypesize{\small}
\tablecaption{Example parameters for steady state evaporation and condensation fronts in the ISM 
(calculated from IIK06), assuming B = 6 $\mu$ G}
\tablewidth{0pt}
\tablehead{
  \colhead{Front Type} & \colhead{P (K cm$^{-3}$)} &
  \colhead{$T_U$ (K)} & \colhead{n$_U$ (cm$^{-3}$)} &
  \colhead{$r_d$} & \colhead{$v_U$ (km \hspace{0.05 cm} s$^{-1}$)} &
  \colhead{A}
}
\startdata
{\bf Evaporation} & 1200 & 35.5 & 33.8 & 0.004 & 0.14 & 0.95 \\
{\bf Condensation} & 4000 & 8115 & 0.49 & 371 & 0.0302 & 11,900\\
\tableline
\enddata  
\end{deluxetable}

\end{document}